\documentclass[sigconf]{acmart}
\AtBeginDocument{%
  \providecommand\BibTeX{{%
    \normalfont B\kern-0.5em{\scshape i\kern-0.25em b}\kern-0.8em\TeX}}}

\copyrightyear{2021}
\acmYear{2021}
\setcopyright{acmcopyright}\acmConference[WI-IAT '21]{IEEE/WIC/ACM
International Conference on Web Intelligence}{December 14--17,
2021}{ESSENDON, VIC, Australia}
\acmBooktitle{IEEE/WIC/ACM International Conference on Web Intelligence
(WI-IAT '21), December 14--17, 2021, ESSENDON, VIC, Australia}
\acmPrice{15.00}
\acmDOI{10.1145/3486622.3493922}
\acmISBN{978-1-4503-9115-3/21/12}

\usepackage{optidef}

\begin{document}

\title{Redistribution in Public Project Problems via Neural Networks}

\author{Guanhua Wang}
\affiliation{%
  \institution{The University of Adelaide}
  \city{Adelaide}
  \state{South Australia}
  \country{Australia}
  \postcode{5005}
}
\email{guanhua.wang@adelaide.edu.au}
\author{Wuli Zuo}
\affiliation{%
  \institution{The University of Adelaide}
  \city{Adelaide}
  \state{South Australia}
  \country{Australia}
  \postcode{5005}
}
\email{wuli.zuo@student.adelaide.edu.au}
\author{Mingyu Guo}

\affiliation{%
  \institution{The University of Adelaide}
  \city{Adelaide}
  \state{South Australia}
  \country{Australia}
  \postcode{5005}
}
\email{mingyu.guo@adelaide.edu.au}

\renewcommand{\shortauthors}{Wang et al.}

\begin{abstract}
Many important problems in multiagent systems involve resource allocations. Self-interested agents may lie about their valuations if doing so increases their own utilities. Therefore, it is necessary to design mechanisms (collective decision-making rules) with desired properties and objectives. The VCG redistribution mechanisms are efficient (the agents who value the resources the most will be allocated), strategy-proof (the agents have no incentives to lie about their valuations), and weakly budget-balanced (no deficits).

We focus on the VCG redistribution mechanisms for the classic public project problem, where a group of agents needs to decide whether or not to build a non-excludable public project. We design mechanisms via neural networks with two welfare-maximizing objectives: optimal in the worst case and optimal in expectation. Previous studies showed two worst-case optimal mechanisms for 3 agents, but worst-case optimal mechanisms have not been identified for more than 3 agents. For maximizing expected welfare, there are no existing results. 

We use neural networks to design VCG redistribution mechanisms.
Neural networks have been used to design the redistribution mechanisms for multi-unit auctions with unit demand.
We show that for the public project problem, the previously proposed neural networks, which led to optimal/near-optimal mechanisms for multi-unit auctions with unit demand, perform abysmally for the public project problem. We significantly improve the existing networks on multiple fronts:
We conduct a GAN network to generate worst-case type profiles and feed prior distribution into loss function to provide quality gradients for the optimal-in-expectation objective. We adopt dimension reduction to handle a larger number of agents and we adopt supervised learning into the best manual mechanism as initialization, then leave it into unsupervised learning. For the worst case, we get better results than the existing manual mechanisms, and for the optimal-in-expectation objective, our mechanisms' performances are close to the theoretical optimal performance.
\end{abstract}




\maketitle

\section{Introduction}

\subsection{VCG Redistribution Mechanisms}

Many important problems in multiagent systems are related to resource allocations. The problem of allocating one or more resources among a group of competing agents can be solved through economic allocation mechanisms that take the agents’ reported valuations for the resources as input, and produce an allocation of the resources to the agents, as well as payments to be made by the agents. As a central research branch in economics and game theory, mechanism design concerns designing collective decision-making rules for multiple agents, to achieve desirable objectives, such as maximizing the social welfare, while each agent pursues her own utility. A mechanism is efficient if the agents who value the resource the most will get it. A mechanism is strategy-proof if the agents have the incentives to report their valuations truthfully, which is to say, an agent's utility is maximized when reporting her true valuation, no matter how the other agents report.

The Vickrey-Clarke-Groves (VCG) mechanism is a celebrated efficient and strategy-proof mechanism. Under the VCG mechanism, each agent $i$ reports her private type $\theta_i$. The outcome that maximizes the agents' total valuations is chosen. Every agent is required to make a VCG payment $t(\theta_{-i})$, which is determined by the other agents' types. An agent's VCG payment is often described as how much this agent's presence hurts the other agents, in terms of the other agents' total valuations. The total VCG payment may be quite large, leading to decreased welfare for the agents. In particular, in the context of the public project problem, where the goal is often to maximize the social welfare (the agents' total utility considering payments), having large VCG payments are undesirable.


To address the welfare loss due to the VCG payments, Cavallo \cite{Cavallo2006:Optimal} suggested that we first execute the VCG mechanism and then redistribute as much of the payments back to the agents, without violating the efficiency and strategy-proofness of the VCG mechanism, and in a weakly budget-balanced way. This is referred to as the VCG redistribution mechanism. The amount that every agent receives (or pays additionally) is called the redistribution payment. To maintain efficiency and strategy-proofness of VCG, the redistribution payment of an agent is required to be independent of her own valuation. To maintain weakly budget-balance, the total amount redistributed should never exceed the total VCG payment. The redistribution payment is characterized by a redistribution function $h$, where
$h(\theta_{-i})$ represents agent $i$'s redistribution payment.

There have been many successes on designing redistribution mechanisms for various multi-unit/combinatorial auction settings \cite{Cavallo2006:Optimal, Clippel2014:Destroy, Faltings2005:Budget-Balanced, Guo2009:Worst, Moulin2009:Almost, Gujar2011:Redistribution, Guo2011:VCG, Guo2012:Worst, Guo2014:Better}, including a long list of optimal/near-optimal mechanisms.
On the other hand, there hasn't been comparable success in solving for optimal redistribution mechanisms for the public project problem, despite multiple attempts \cite{Naroditskiy2012:Redistribution, Guo2016:Competitive, Guo2017:Speed, Guo2019:Asymptotically}. In terms of optimal results, Naroditskiy {\em et al.} \cite{Naroditskiy2012:Redistribution} solved for the worst-case optimal mechanism for
three agents. Unfortunately, the authors' technique does not generalize to more than three agents. Guo \cite{Guo2019:Asymptotically} proposed a mechanism that is worst-case optimal when the number of agents approaches infinity, but for small number of agents, the mechanism is not optimal. For maximizing expected welfare, there are no existing results, because it is difficult for traditional mathematical analysis(eg: mixed integer programming) to maximize the expectation of welfare. 

\subsection{Designing VCG Redistribution Mechanisms via Neural Networks}

A recent emerging topic in mechanism design is to bring tools such as neural networks from machine learning to design mechanisms \cite{Duetting2019:Optimal, Golowich2018:Deep, Manisha2018:Learning, Shen2019:Automated, Wang2021:Mechanism}. Duetting et al. \cite{Duetting2019:Optimal} proposed a neural network approach for the automated design of optimal auctions. They model an auction as a multi-layer neural network and frame optimal auction design as a constrained learning problem which can be solved using standard machine learning pipelines. {\em The training and testing type profiles are generated based on the prior distribution. The cost function involves the mechanism objective and the penalty for property violation.} Essentially, neural networks were used as tools for functional optimisation. Shen et al. \cite{Shen2019:Automated} proposed a neural network based framework to automatically design revenue optimal mechanisms. This framework consists of a seller’s network, which provides a menu of options to the buyers, and a buyer’s network, which outputs an action that maximizes her utility. Wang et al. \cite{Wang2021:Mechanism} studied mechanism design for the public project problem and proposed several technical innovations that can be applied to mechanism design in general to improve the performance of mechanism design via neural networks. 

\begin{figure}[h]
  \centering
  \includegraphics[width=0.6\linewidth,height=5.5cm]{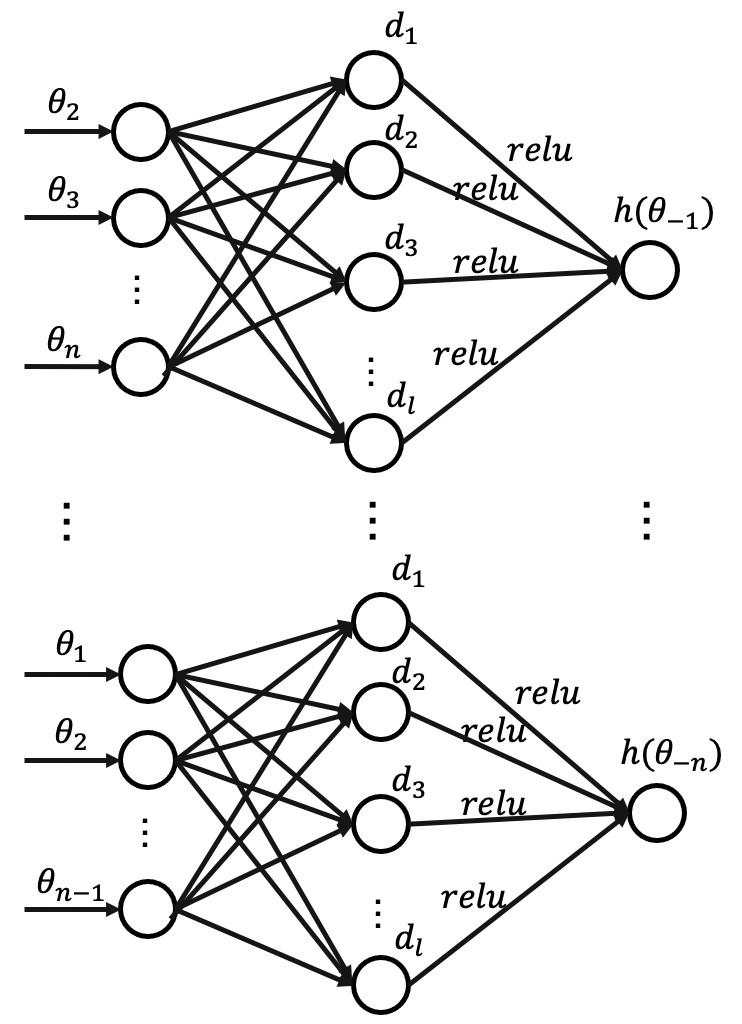}
  \caption{Neural network structure reported by Manisha et al. \cite{Manisha2018:Learning}}
  \label{fig:ManishaNet}
\end{figure}

The work by Manisha et al. \cite{Manisha2018:Learning} is the first and only attempt to design VCG redistribution mechanisms using neural networks. They focused on multi-unit auctions with unit demand and studied both worst-case and optimal-in-expectation objectives. By randomly generating a large number of bid profiles, they train a neural network to maximize the total redistribution payment, while enforcing that the total redistribution should not exceed the total VCG payment. They modelled the redistribution function as a simple network outlined in Figure \ref{fig:ManishaNet}. It is a fully connected network with one hidden layer using ReLU activation. It takes the valuations of all the agents other than agent $i$ herself as input, and outputs the predicted redistribution payment for agent $i$. Their data is sampled randomly from uniform distribution ($\theta_i \in Uniform(0, 1)$).


\subsection{Improved Neural Networks for Designing VCG Redistribution Mechanisms for the Public Project Problem}

In this paper, we train neural networks to design VCG redistribution functions for the public project problem, which turns out to be a more challenging setting compared to multi-unit auctions studied by Manisha et al. \cite{Manisha2018:Learning}. The public project problem is a classic mechanism design problem that has been studied extensively in economics and computer science \cite{Mas-Colell1995:Microeconomic, Moore2006:General, Moulin1988:Axioms}. In this problem, $n$ agents decide whether or not to build a non-excludable public project, for example, a public bridge that can be accessed by everyone once built. Without loss of generality, we assume that the cost of the project is $1$, and $\theta_{i} (0\leq\theta_{i}\leq1)$ is agent $i$’s valuation for the project if it is built. If the decision is not to build, every agent retains her share of the cost, which is $1/n$.




We first evaluate the simple multilayer perceptron (MLP) model 
proposed by Manisha et al. \cite{Manisha2018:Learning}. 
That is, for each agent $i$, we train a neural network that maps $\theta_{-i}$ to agent $i$'s redistribution. The training and testing samples are randomly generated based on the prior distribution.
The cost function maximizes the mechanism design objective, as well as enforces mechanism design constraints via penalty.
We find that such a simple MLP is not effective enough for the public project problem for the following reasons:

\begin{enumerate}
  \item From our experiments, by randomly generating the type profiles, 
  we are not getting the true worst-case type profiles for the public project problem (it is a coincidence that for multi-unit auctions with unit demand, it is a lot easier to hit a worst case).
  \item Another challenge is the high input dimension when the number of agents is large. For 100 agents, the neural network has to take a 99-dimensional input, which is computationally unrealistic.
  \item In the public project problem, the agents' collective payments differ significantly between cases where the decision is to build the public project and cases
  where the decision is not to build.
  The number of ``build'' samples in a training batch significantly impacts
  the parameter gradient, which results in a wild loss fluctuation during the training process.
\end{enumerate}

To solve the aforementioned problems, we propose a novel neural network approach to design redistribution mechanisms for the public project problems. 
Our approach involves the following technical innovations.

\paragraph{GAN Network} For the worst-case objective, we introduce a generative adversarial network (GAN) to generate worst-case type profiles, and then use these type profiles to train the redistribution function. Our experiment shows that a mechanism trained only using randomly generated data fails when facing 
type profiles generated by GAN, so GAN is necessary and effective to derive the worst case.

\paragraph{Dimension Reduction} Instead of feeding $\theta_{-i}$ as input to train the mechanism, which has $n-1$ dimensions, we extract a few expressive features from $\theta_{-i}$ (e.g. the maximum of types $\theta_{-i}$, the sum of $\theta_{-i}$ excluding the maximum, etc.). This reduces the input dimension to 3. This helps the neural network loss converge faster and still retain good performance.

\paragraph{Supervised Learning} Wang et al. \cite{Wang2021:Mechanism} suggested that supervision to manual mechanisms often outperforms random initialization in terms of training speed by pushing the performance to a state that is already somewhat close to optimality. In addition, unlike many other deep learning problems, for mechanism design, there often exist simple and well-performing mechanisms that can be used as starting points. In this particular problem, we first conduct supervised learning to let the network mimic the state of art manual mechanism \cite{Guo2019:Asymptotically}, and then leave it to gradient descent. This approach saves time for larger $n$ in our experiments.

\paragraph{Feeding Prior Distribution into Loss Function} We use probability density function (PDF) of the prior distribution to provide quality gradients. In specific, for each valuation profile $\theta = \{\theta_i\} (i=1..n)$ generated from a distribution $D$, we randomly choose $\theta_i$ to be replaced by a randomly generated $\theta'_i$ from $Uniform(0,1)$ and update $\theta$ to be $\theta'$. This sample is then assigned a weight based on the PDF. 
In experiments, we see that this approach significantly reduces the loss fluctuation during training. One potential explanation (or observation) is that this approach reduces the fluctuation in the proportion of ``build'' cases among a batch.

With a more sophisticated network architecture due to the above technical adjustments, we get better results for the worst-case than the state of the art \cite{Guo2019:Asymptotically}. For the optimal-in-expectation objective, our results are close to the theoretical optimal values.

\subsection{Organization}

The rest of the paper is organized as followings: Section 2 gives a formal description of the VCG redistribution mechanism for the public project problem. Section 3 and 4 present our neural network approaches for the worst-case and expectation optimal objectives respectively, with the explanation of the network architecture, the technical adjustments, and loss function design. Section 5 discusses the training of neural networks and the experimental analysis. Section 6 concludes.

\section{Formal Model Description}

For the public project problem, VCG redistribution mechanisms have the following form \cite{Naroditskiy2012:Redistribution}: 
\begin{itemize}
  \item	Build the public project if and only if $\sum_i\theta_{i}\geq1$.
  \item	If the decision is to build, agent i receives $\sum_{j\neq i}\theta_{j}-h(\theta_{-i})$.
  \item	If the decision is not to build, then agent i receives $(n-1)/n-h(\theta_{-i})$.
  \item	$h$ is an arbitrary function and $\theta_{-i}$ refers to the types from the agents other than $i$ herself.
\end{itemize}

A VCG redistribution mechanism is characterized by the function $h$.

Due to Guo \cite{Guo2019:Asymptotically}, 

\begin{itemize}
  \item $S(\theta)=max\{\sum_i \theta_i, 1\}$ is exactly the first-best total utility. (I.e., if the sum of types is higher than $1$, then the efficient decision is to build. Otherwise, the efficient decision is not to build.)
  \item The agents' welfare (total utility considering payments) under type profile $\theta$ is $nS(\theta)-\sum_i h(\theta_{-i})$, which is obtained via simple algebraic simplification based on the definition of VCG redistribution mechanisms.
\end{itemize}

We consider two objectives. One is to find a mechanism that maximizes the worst-case efficiency ratio, and the other is to maximize the expected efficiency ratio.

\subsection{Worst-case Optimal Mechanism}

The efficiency ratio $r$ is defined as the ratio between the achieved total utility and the first best total utility:
$$r = \frac{nS(\theta)-\sum_i{h(\theta_{-i})}}{S(\theta)} = n - \frac{\sum_i{h(\theta_{-i})}}{S(\theta)}$$

The worst-case efficiency ratio is the worst case ratio between the achieved total utility and the first best total utility, namely, the minimum of $r$ over all type profiles. 
Due to Guo \cite{Guo2019:Asymptotically}, the mechanism has a worst-case efficiency ratio $\alpha$ if and only if:
    \begin{equation}\label{equ:worst-const}
      \forall\theta, (n-1)\leq\sum_i h(\theta_{-i})/S(\theta)\leq(n-\alpha)
    \end{equation}

In Inequality \ref{equ:worst-const}, the left side is the constraint for weakly budget-balance, and the right side corresponds to the definition of $\alpha$.

Therefore, taking the worst-case ratio as the objective, we need to design an $h$ function that:

\begin{maxi}[3]
{}{\alpha}{}{}
\addConstraint{\forall{\theta},(n-1)}{ \leq \sum_i{h(\theta_{-i})}/S(\theta)}{\leq (n-\alpha)}
\end{maxi}

\subsection{Optimal-in-Expectation Mechanism}

For this objective, we maximize the expected efficiency ratio $r$ to 1, which is equivalent to minimize $\sum_i{h(\theta_{-i})}/S(\theta)$ from above to $n-1$, with the consideration of the weakly budget-balance constraint.

We are designing an $h$ function that:

\begin{mini}[3]
{}{\overline {\sum_i{h(\theta_{-i})}/S(\theta)}}{}{}
\addConstraint{\forall{\theta},(n-1)}{\leq \sum_i{h(\theta_{-i})}/S(\theta)}
\end{mini}

\section{Worst-case Optimal Mechanism}

In this section, we focus on the worst-case optimal mechanism. We first describe our neural network approach by explaining the network architecture and details of the relating techniques. Then we define the loss function.  

\subsection{Network Architecture}

\begin{figure*}[htbp]
\centering
\includegraphics[width=0.9\textwidth,height=6cm]{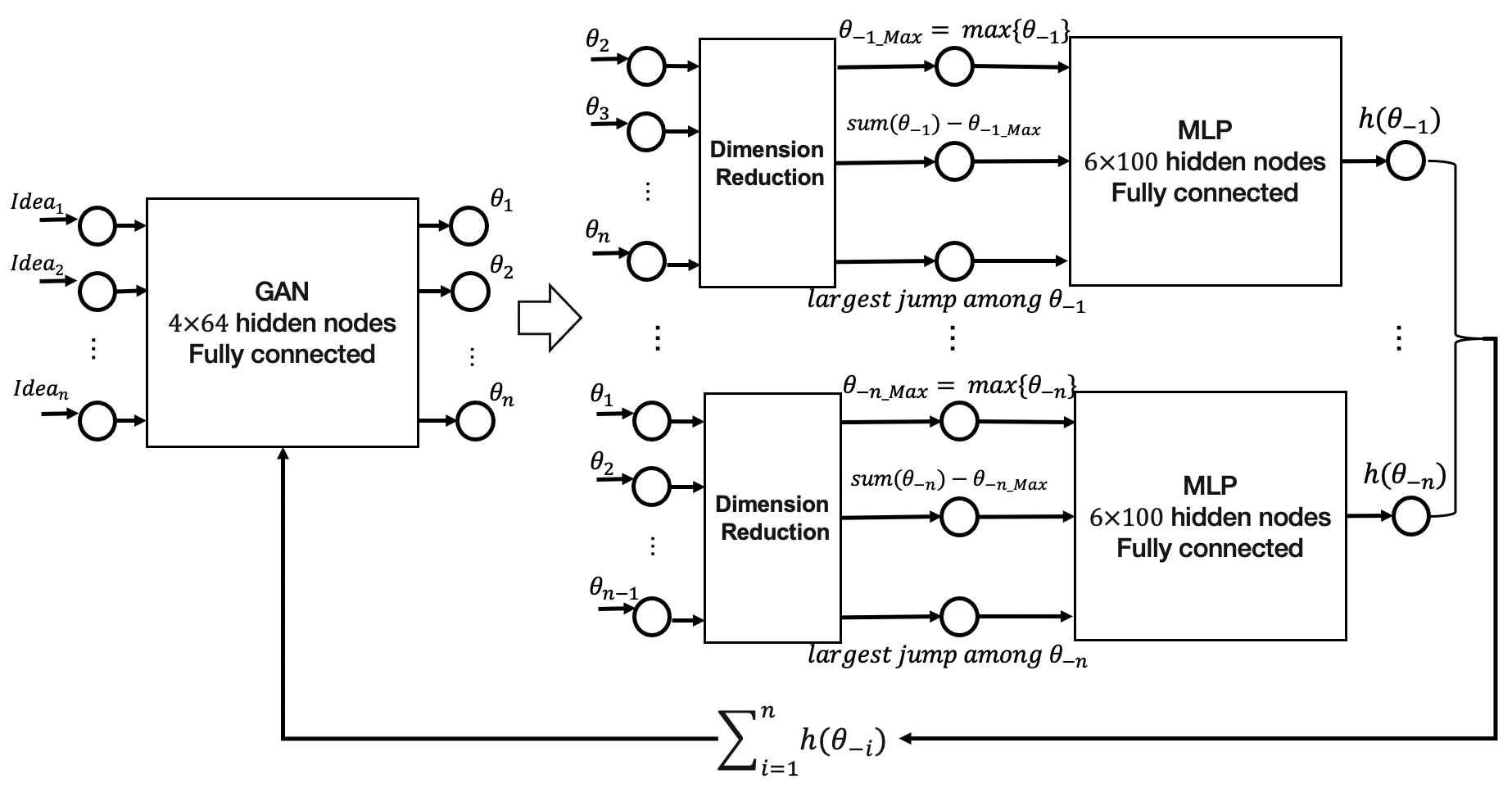}
\caption{Worst-case optimal network architecture}
\label{fig:GAN-MLP}
\end{figure*}

We construct a neural network to determine the $h$ function. As illustrated in Figure \ref{fig:GAN-MLP}, it is a network system in which a GAN and an MLP interacting with each other, we call it GAN+MLP. 

The GAN works as a Generative Model and is used to generate special samples (type profiles). It takes $n$ randomly generated values as input ideas, and the output is $\theta=\{\theta_i\} (i=1...n)$. It is a fully connected network with 4 hidden layers, and each hidden layer contains 64 nodes. For a given batch ($batch\_size=b$), the GAN generates $b$ type profiles: $\theta^{(j)}\in batch = \{\theta^{(1)}, \theta^{(2)},...,\theta^{(b)}\}$, with the aim to maximize the difference between the maximum and the minimum of $\sum_i{h(\theta^{(j)}_{-i})}/S(\theta^{(j)})$. It means to:

$$maximize\ (\sum_i {h(\theta^{(j_1)}_{-i})}/S(\theta^{(j_1)}) - \sum_i{h(\theta^{(j_2)}_{-i})}/S(\theta^{(j_2)}))$$
where $\theta^{(j1)}, \theta^{(j2)}\in batch$ is the sample that gives the maximum and minimum of $\sum_i{h(\theta^{(j)}_{-i})}/S(\theta^{(j)})$, respectively.

The MLP works as a Discriminative Model that learns the samples generated by the GAN. The MLP is a fully connected neural network. For each agent $i$, the network takes  $\theta_{-i}$ as the input, and outputs the value of $h(\theta_{-i})$. In the MLP, there are 6 hidden layers, each of which contains 100 nodes and with ReLU as the activation function. We first train the MLP under supervision to the best-performing manual mechanism and then leave it to unsupervised learning. For unsupervised learning, our cost function is the combination of design objective and also penalty due to constraint violation.

\subsection{Details of the Networks and Evaluations}

To improve the result for the neural network, we adopt some technical adjustments. We use a GAN instead of uniform to generate special cases in order to find out the worst case. For cases with a greater number of agents ($n\geq 5$), we adopt two technical tricks: Dimension Reduction and Supervised Learning. 

\subsubsection{GAN Network}

In previous studies, the authors used random generation or fixed data to find the worst case \cite{Manisha2018:Learning}. 

As mentioned in Section 3.1, We propose a new GAN approach to find out the worst case.

We conduct a contrast experiment to verify the validation of the GAN. We use only data generated from uniform distribution to train a network, and test this network with two sets of data.

\begin{itemize}
    \item Test Set A: 20000 data drawn from $Uniform(0, 1)$
    \item Test Set B: 10000 data generated from a trained GAN network and 10000 data drawn from $Uniform(0, 1)$
\end{itemize}

\begin{figure}[h]
\begin{minipage}{0.23\textwidth}
 \centering
 \includegraphics[width=\linewidth]{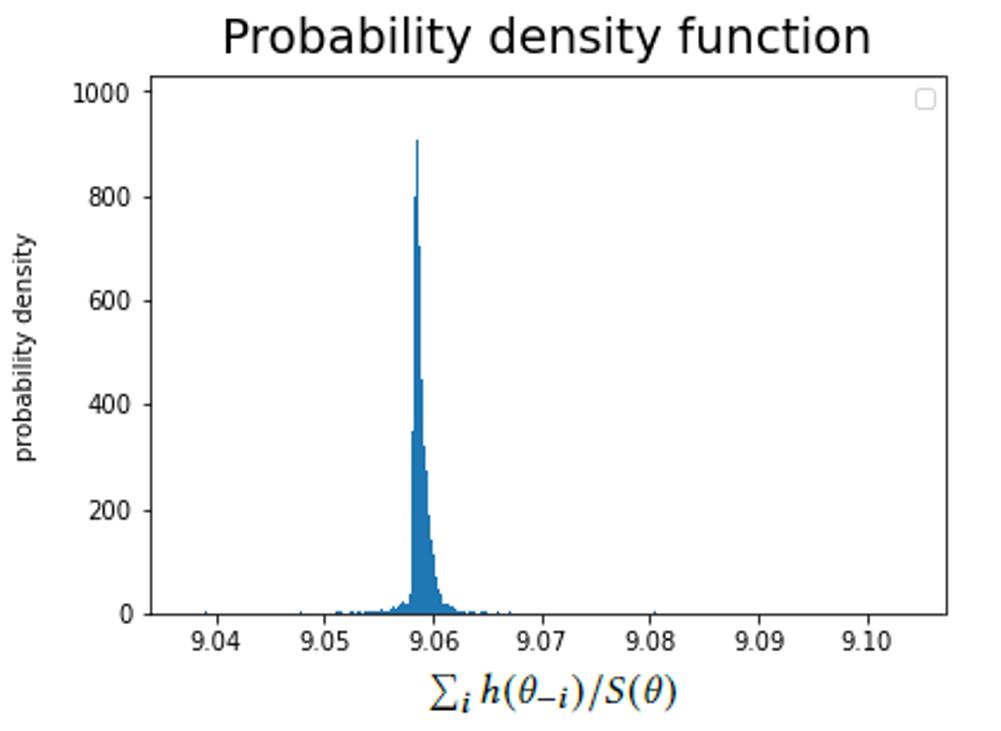}
\end{minipage}\hfill
\begin {minipage}{0.23\textwidth}
 \centering
 \includegraphics[width=\linewidth]{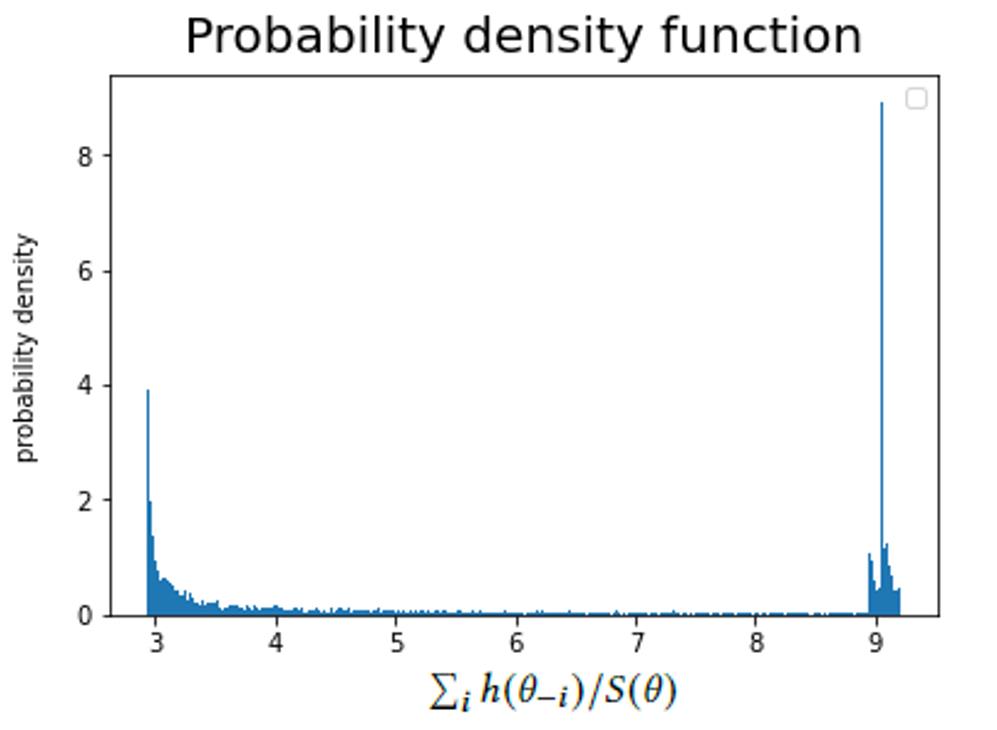}
\end{minipage}
\caption{Spread of $\sum_{i}{h(\theta_{-i})} / S(\theta)$ evaluated using random type profiles and GAN generated type profiles. Random generation fails to generate true worst cases.} \label{fig:GAN-comp}
\end{figure}

Figure \ref{fig:GAN-comp} outlines the experimental results for $n=10$ showing the difference of the the network performance on Test Set A and B. In the left figure, the network is tested by randomly generated data set A. It gives $\alpha=0.896$, and $\sum_{i}{h(\theta_{-i})}/ S(\theta)$ is between 9 and 9.104. However, in the right figure, the network performs poorly with significant violation of the weakly budget-balance constraint ($\sum_{i}{h(\theta_{-i})} / S(\theta)$ is from 3 to 9.2). 

Therefore, random type profile generation as used in Manisha et al. \cite{Manisha2018:Learning} does not work for this problem. To get the worst-case performance, we need a GAN network to generate higher quality worst-case profiles, and then let the MLP learn these profiles.

\subsubsection{Dimension Reduction}

The MLP takes a $(n-1)$-dimensional input for $n$ agents, resulting in expensive computation when $n$ is large. This motivates us to look for an effective dimension-reducing technique.

We first manually design a list of features that describe $\theta_{-i}$, and then experimentally search for a good combination of three features to be used for dimension reduction purposes. (We essentially reduce $\theta_{-i}$ to three dimensions this way.)


The features we consider include:
\begin{itemize}
  \item The highest type(s) from $\theta_{-i}$
  \item The the lowest type(s) from $\theta_{-i}$
  \item The sum of some types from $\theta_{-i}$
  \item The standard deviation of some types from $\theta_{-i}$
  \item The largest jump of adjacent types from $\theta_{-i}$\\
  Here $jump\ of\ adjacent\ types$ is defined as: for a sorted list ${\theta_{-i}}$, there is $jump_j$ between $\theta_j$ and $\theta_{j+1}$: 
  $$jump_j = \theta_{j+1}-\theta_j$$
\end{itemize}


We first experimentally derive that the following features are more important than the rest (i.e., removing them results in significant performance loss): 

\begin{itemize}
  \item The highest type from $\theta_{-i}$
  \item The the lowest type from $\theta_{-i}$
  \item The sum of some type from $\theta_{-i}$
\end{itemize}

We then experimentally evaluate different combinations of the above features:

\begin{enumerate}
\item   Combination 1: the highest type(s) from $\theta_{-i}$ \& the sum of all the other types
\item   Combination 2: the highest type(s) from $\theta_{-i}$ \& the difference between the highest and the lowest type from $\theta_{-i}$
\item   Combination 3: the highest type(s) from $\theta_{-i}$ \& the standard deviation of all types from $\theta_{-i}$
\item   Combination 4: the highest type(s) from $\theta_{-i}$ \& the standard deviation of all the other types
\item   Combination 5: the highest type(s) from $\theta_{-i}$ \& the largest jump of adjacent types
\item   Combination 6: the highest type(s) from $\theta_{-i}$, the lowest type from $\theta_{-i}$, \& the largest jump of adjacent types
\item   Combination 7: the highest type(s) from $\theta_{-i}$, the sum of all the other types \& the largest jump of adjacent types
\item   Combination 8: the highest type(s) from $\theta_{-i}$, the lowest type from $\theta_{-i}$, \& the sum of all the other types
\end{enumerate}

\begin{figure}[h]
  \centering
  \includegraphics[width=0.72\linewidth]{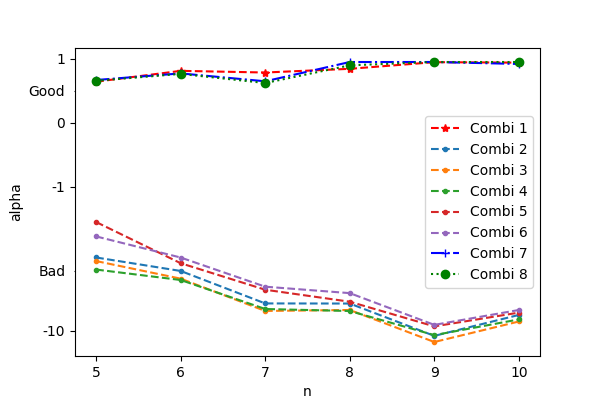}
  \caption{The worst-case ratio $\alpha$ of the models trained with the input generated by using different feature-combinations against the number of agents $n=5, ..., 10$.
}
  \label{fig:CombiComp-alpha}
  \Description{The worst-case ratio $\alpha$ of the models trained with the input generated by using different feature-combinations against the number of agents $n=5, ..., 10$.}
\end{figure}

Figure \ref{fig:CombiComp-alpha} shows $\alpha$ of the networks trained with the different input combinations against the number of agents $n$. It is found that Combination 1, 7 and 8 performs better than the other combinations.

The above dimension-reducing mechanism improves both the training speed and sometimes the performance. We can infer that by using this dimension-reducing mechanism, the training speed would have a more significant improvement with the increase of the agent number $n$.

\begin{figure}[h]
  \centering
  \includegraphics[width=0.9\linewidth]{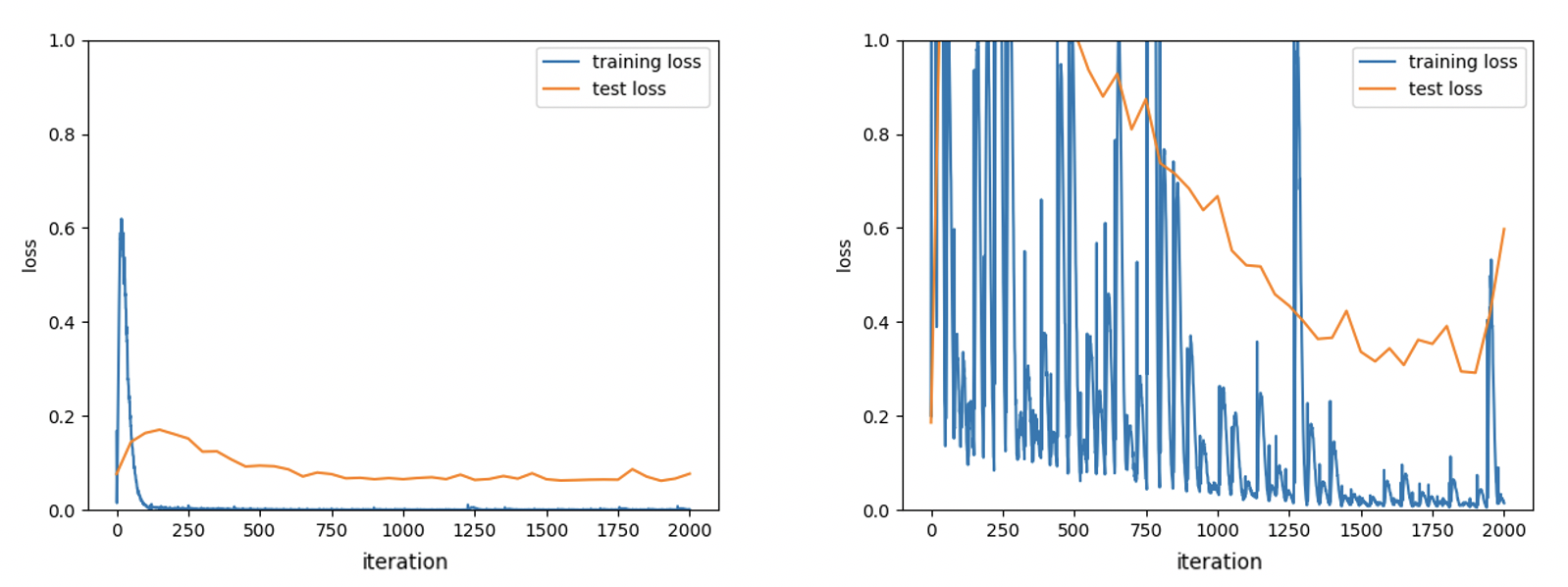}
  \caption{The loss during the training when using (left) and not using (right) the dimension-reducing mechanism for $n=10\ (\theta_i \in Uniform(0,1))$. Dimension reduction leads to faster convergence.}
  \label{fig:comp-loss}
\end{figure}

Figure \ref{fig:comp-loss} shows the difference of the loss of the network between using and not using dimension-reducing mechanism for $n=10$. In the left figure, the training and test loss is stabilized within 1000 iterations with the application of the dimension-reducing mechanism, while in the right figure, the losses still vibrate in a wider range till 2000 iterations.

\begin{figure}[h]
  \centering
  \includegraphics[width=\linewidth]{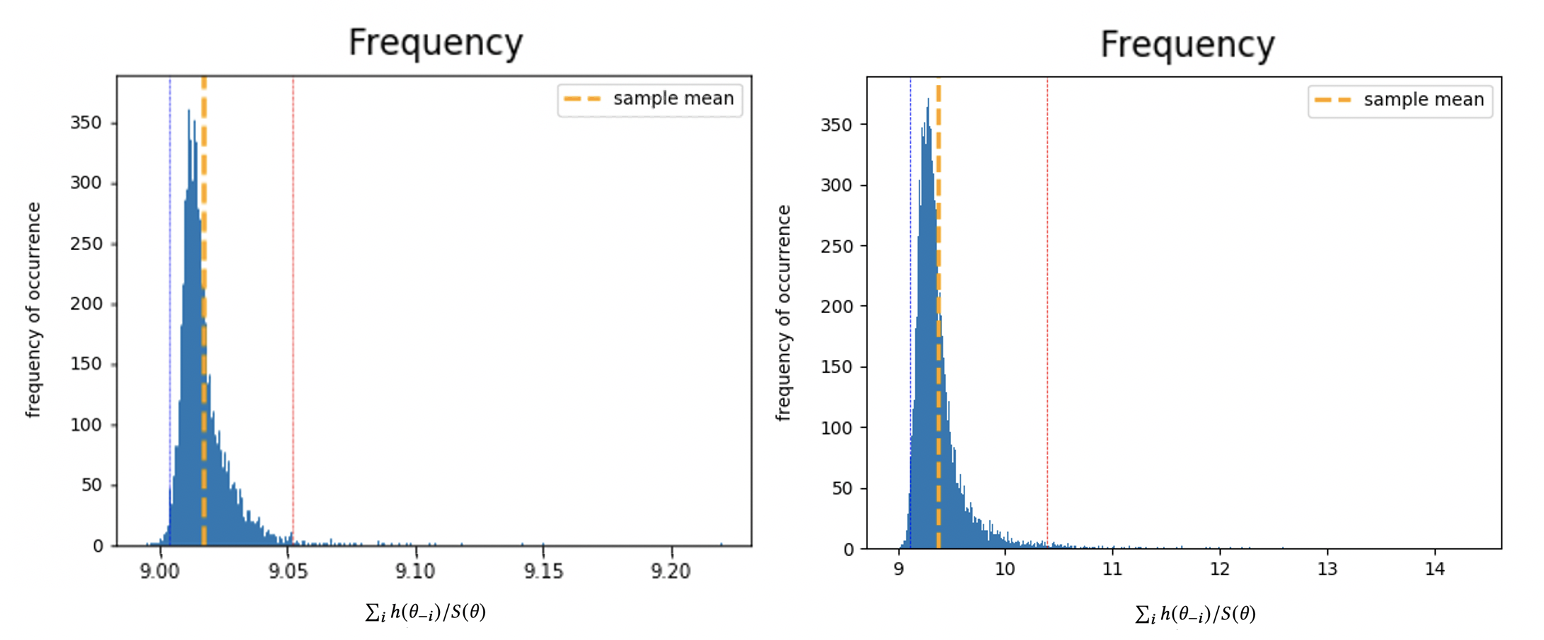}
  \caption{Spreading of $\sum_i h(\theta_{-i})/S(\theta)$ when using (left) and not using (right) the dimension-reducing mechanism for $n=10\ (\theta_i \in Uniform(0,1))$, data size = 10000, bins = 500. Dimension reduction leads to better-performing mechanism.}
  \label{fig:comp-sumh-S}
\end{figure}


Figure \ref{fig:comp-sumh-S} shows the difference on the spreading of $\sum_i{ h(\theta_{-i})}/S(\theta)$ between using and not using dimension-reducing mechanism for $n=10$. In the left figure, when using the dimension-reducing mechanism, we get an expected performance of $\sum_i{h(\theta_{-i})}/S(\theta) = 9.003$, which is very close to $n-1=9$. This indicates the corresponding mechanism is close to being optimal. (We recall that for the expected performance, we want the expected ratio to be as close to $n-1$ as possible.)
When not using the dimension-reducing mechanism, our performance is $9.377$, which is further away from $n-1=9$.

\subsubsection{Supervised Learning}

For a greater number of agents, it takes gradient descent a long time to fix the constraint violations. Wang et al. \cite{Wang2021:Mechanism} suggested that we first supervise the neural network into the existing best manual mechanism, and then leave it to unsupervised learning. With the best manual mechanism as the starting point, better mechanisms can usually be found.
The existing best manual mechanism for worst-case optimal objective is reported by Guo \cite{Guo2019:Asymptotically}.



\subsection{Loss Function}

\subsubsection{GAN Network Loss}
The GAN network has the following loss:
\begin{align*}
   loss_{GAN} = & min\{\sum_i { h(\theta^{(j_1)}_{-i})}/S(\theta^{(j_1)})\}\\
   & - max\{\sum_i{h(\theta^{(j_2)}_{-i})}/S(\theta^{(j_2)})\},
   \ & \theta^{(j_1)},\theta^{(j_2)} \in batch
\end{align*}

\subsubsection{Supervised Loss}

In supervised learning, we want the predicted $h$ to be as close as possible to the best manual value $h\_manual$ \cite{Guo2019:Asymptotically}, so the loss is:
$$loss_{supervised} = (h - h\_manual)^2$$

\subsubsection{Unsupervised Loss}

In the unsupervised learning stage, we want all the $\sum_{i}{h(\theta_{-i})}/S(\theta)$ to maintain the weakly budget-balance constraint, and let $\alpha$ to be close to 1, which means to make the upper bound of $\sum_{i}{h(\theta_{-i})}/S(\theta)$ as small as possible.

The loss function includes two parts:

\begin{itemize}
\item $objective\_loss$ = $(relu(\sum_{i}{h(\theta_{-i})} - (n-up\_bound) S(\theta)))^2 $
\item $constraint\_loss$ = $(relu((n-1)S(\theta)- \sum_{i}{h(\theta_{-i})}))^2$
\end{itemize}


Since the weakly budget-balance is a strict constraint, while the objective is soft, we add a multiplier $\epsilon$ to weaken the effect of $objective\_loss$. For the worst-case optimal network, we get the best $\epsilon=0.01$ through experiments.
\begin{align*}
loss_{unsupervised} &= \epsilon \cdot objective\_loss + constraint\_loss\\
&= objective\_loss/100 + constraint\_loss
\end{align*}

\section{Optimal-in-Expectation Mechanism}

We design the optimal-in-expectation mechanism with slight modifications based on the worst-case mechanism.

The architecture of the MLP stays the same. We do not need a GAN for generating the worst-case since the worst-case does not matter. For a large agent number $n$, we also adopt dimension reduction and supervised learning as we do for the worst-case mechanism. In addition, we feed the prior distribution into the loss function to achieve a high-quality gradient. This network is called MLP+FEED.

\subsection{Feed Prior Distribution into Loss Function}

In this problem, the decision to build or not to build significantly affects the expectation of a training batch. For batches with different amounts of ``build'' cases, the gradients fluctuate significantly, causing worse training results and speed.

Wang et al. \cite{Wang2021:Mechanism} discovered a way to insert the cumulative distribution function (CDF) of the prior distribution into the cost function to get more accurate loss function. The approach was shown to be effective for optimal-in-expecation mechanism design. We adopt a similar idea, but use probability density function (PDF) from the prior distribution to provide quality gradients for our training process. For each valuation profile $\theta = \{\theta_i\} (i=1..n)$ generated from a distribution $D$, we randomly choose $\theta_i$ to be replaced by $\theta'_i$ which is regenerated from $Uniform(0,1)$ and update $\theta$ to be $\theta'$. 

The probability of the profile $\theta'$ is proportional to $PDF_D(\theta'_i)$, so the loss should also be multiplied by $PDF_D(\theta'_i$).
$$loss_{unsupervised\_feeding} = loss_{unsupervised} \cdot PDF_D(\theta'_i)$$

Here, 
$$PDF_D(\theta'_i) = 10^{log\_prob(\theta'_i)}$$

\begin{figure}[h]
  \centering
  \includegraphics[width=0.7\linewidth]{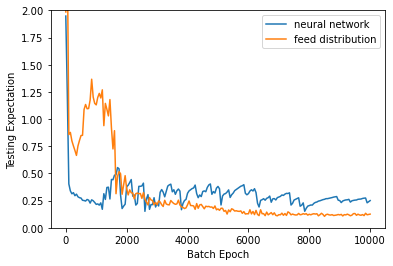}
  \caption{Test loss (the distance from $\sum_{i}{h(\theta_{-i})} / S(\theta)$ to $(n-1)$) during the training for distribution: $Normal(0.5,0.1),\ n=3)$. Feeding prior distribution leads to faster convergence and better mechanism.} \label{fig:Feed}
\end{figure}

$log\_prob$ is provided by PyTorch to calculate PDF. PyTorch distributions package is based on Schulman \cite{schulman2015gradient}. 
Our experiments show the difference between feeding and not feeding the distribution into loss function. Figure \ref{fig:Feed} shows that for normal distribution ($Normal(0.5,0.1),\ n=3$), feeding distribution into the loss function helps get a better gradient and thus dramatically improves the test result. 


\subsection{Loss Function}

The loss for the optimal-in-expectation network is similar to that of the worst-case MLP.

\subsubsection{Supervised Loss}

$$loss_{supervised} = (h - h\_manual)^2 $$

\subsubsection{Unsupervised Loss}

We use the square loss to approximate the loss for the expectation, since the derivative of square loss ($f(x) =x^2$) is a linear function, and can represent the strength of the gradient decedent, which is also linear ($f'(x)=ax+b$). 

Loss function consists of $objective\_loss$ and $constraint\_loss$, where

\begin{itemize}
\item $objective\_loss = (relu(\sum_{i}{h(\theta_{-i})} - (n-1)S(\theta)))^2$ 
\item $constraint\_loss = (relu((n-1) S(\theta)- \sum_{i}{h(\theta_{-i})}))^2$
\end{itemize}

$objective\_loss$ aims to push $\sum_{i}{h(\theta_{-i})}/S(\theta)$ close to $(n-1)$, so that the mechanism redistributes as much of the collected VCG payment back to the agents. $constraint\_loss$ is for weakly budget-balance, to make the total redistributed amount less than the total VCG payment.

Similar with the worst-case optimal network, we use a multiplier $\epsilon$ to soften $objective\_loss$, and the experimentally best $\epsilon=10^{-4}$. So,
\begin{align*}
loss_{unsupervised} &= \epsilon \cdot objective\_loss + constraint\_loss\\
&= objective\_loss/10000 + constraint\_loss
\end{align*}

With the feeding of prior distribution to loss function as described in Section 4.1, we have:
\begin{align*}
&loss_{unsupervised\_feeding}\\
&= (objective\_loss/10000 + constraint\_loss)*PDF_D(\theta'_i)
\end{align*}

\section{Experiments and Results}

We program with python by using the third-party library {\em PyTorch}. The experiments are conducted on a computer with an i5-8300H CPU and an Nvidia 1060 GPU. The experiment running time varies from a few minutes up to about 1 hour, depending on different agent numbers and data sizes.

\subsection{Experiment settings}

\subsubsection{Generate Data}

We randomly generate training and test data from prior distributions, and also use the GAN network. Half of the test data is generated from the GAN and the other half is randomly generated data. The test data size is from 10000 to 100000, and increases with $n$, as shown in Table \ref{tab:data-size}. We generate new test data for each test.

\begin{table}[H]
    \centering
    \small
    \begin{tabular}{cccccccc}
    \hline
        n & 4 & 5 & 6 & 7 & 8 & 9 & 10 \\ \hline
        Data size & 10000 & 20000 & 20000 & 20000 & 50000 & 50000 & 100000 \\
        \hline
    \end{tabular}
    \caption{Test data size for different $n$}
    \label{tab:data-size}
\end{table}

\subsubsection{Batch Size}

Keskar et al.\cite{keskar2017large} show that a larger batch leads to a dramatic degradation in the quality of the model. They investigate the cause for this generalization drop in the large-batch regime and present numerical evidence that supports the view that large-batch methods tend to converge to sharp minimizers of the training and testing functions. It shows that the large batch size converges to the sharp minimum, while the small batch size converges to the flat.

Our experiments support Keskar's view and we find that the loss does not reduce with a big batch size ($\geq 1024$). We set the batch size to 64 through comparative experiments.

\subsubsection{Order input}

Arora et al. \cite{arora2016understanding} shows that it is hard for ReLU to simulate the function $max\{a,b\}$. A single $max\{a,b\}$ needs two layers and 5 nodes with exact weights and biases. So inputting sorted $\theta_{-i}$ values to the MLP is a necessary and important step to get good results.

We order the valuations such that $\theta_1 \geq \theta_2 \geq ... \geq \theta_n$. For redistribution problems, sorted input values will not influence strategy-proofness.


\subsubsection{Initialization and Optimizer}

We use Xavier normal initialization for the weights and $Normal(0,0.01)$ for the bias.

We use Adam optimizer with the $learning\_rate=0.001$ initially, and decays by 0.98 every 100 steps by Pytorch Scheduler.

\subsection{Results}

\subsubsection{Worst-case results}

We compare our result for the worst-case optimal mechanism(GAN+MLP) with the previously proposed mechanisms:
\begin{itemize}
\item SBR: heuristic-based SBR mechanism \cite{Naroditskiy2012:Redistribution}
\item ABR: heuristic-based ABR mechanism \cite{Guo2016:Competitive}
\item AMD: mechanisms derived via Automated Mechanism Design (AMD) \cite{Guo2017:Speed}
\item AO: asymptotically optimal (AO) VCG Redistribution \cite{Guo2019:Asymptotically}
\item UB: the {\bf conjectured} upper bounds (UB) on
the efficiency ratios \cite{Naroditskiy2012:Redistribution}
\end{itemize}

\begin{table}[H]
\centering
\begin{tabular}{ccccccc}
\hline
n & SBR & ABR   & AMD   & AO    & GAN + MLP      & UB    \\ \hline
4  &  0.354      & 0.459 & 0.600 & 0.625 & \textbf{0.634} & 0.666 \\
5  &  0.360       & 0.402 & 0.545 & 0.600 & \textbf{0.622} & 0.714 \\
6  &  0.394       & 0.386 & 0.497 & 0.583 & \textbf{0.592} & 0.868 \\
7  &  n too large       & 0.360 & 0.465 & 0.571 & \textbf{0.626} & 0.748 \\
8  &  n too large         & 0.352 & 0.444 & 0.563 & \textbf{0.654} & 0.755 \\
9   &  n too large        & 0.339 & 0.422 & 0.556 & \textbf{0.682} & 0.772 \\
10  &  n too large        & 0.336 & 0.405 & 0.550 & \textbf{0.623} & 0.882 \\ \hline
\end{tabular}
\caption{GAN+MLP Compare with previous results for worst Case}
\label{tab:Worst-case}
\end{table}

The result in Table \ref{tab:Worst-case} shows that our mechanism achieves better worst-case efficiency ratios than all previous results.

The main downside of our results is that our worst-case is calculated numerically by trying a large number of type profiles. This is a limitation
due to our neural network based approach. (Manisha et al. \cite{Manisha2018:Learning}'s neural network approach also evaluated the worst-case by randomly generating a large number of type profiles. We showed that our GAN approach is a lot more rigorous compared to simply random profile generation.)
It should be noted that SBR, ABR and AMD's worst-cases were also calculated numerically. (AO's worst-case was derived analytically.)
To test the stability of the worst-case ratios, we experimented with different test sizes. Table \ref{tab:data-size} shows that for $n=10$ with different test sizes of 10000, 20000 and 100000, $\alpha$ is stable at $0.623$.

\begin{table}[H]
    \centering
    \begin{tabular}{cccc}
    \hline
        Data size & 10000 & 20000 & 100000 \\\hline
        $\alpha$ & 0.623 & 0.623 & 0.623 \\
        \hline
    \end{tabular}
    \caption{$\alpha$ for Different test size ($n=10$)}
    \label{tab:data-size}
\end{table}


\subsubsection{Optimal-in-expectation results}

We evaluate the expectation of $\sum_{i}{h(\theta_{-i})} / S(\theta)$ under our mechanism obtained via MLP+FEED. As defined in our model, for a case with $n$ agents, the theoretical optimal value for $\sum_{i}{h(\theta_{-i})} / S(\theta)$ is $n-1$, so we want the expectation to be as close to $n-1$ as possible from above (to maintain weakly budget-balance). Our results show that the obtained mechanisms are near optimal. For example, we get $E=4.061$ vs $n-1=4$ for $n=5$, $E=5.034$ vs $n-1=5$ for $n=6$.

Table \ref{tab:Expectation} shows that for both data generated from uniform distribution and normal distribution, the average $\sum_{i}{h(\theta_{-i})}/S(\theta)$ of our MLP+FEED network is very close to the theoretical optimal value $n-1$, which means that the redistribution function will return the vast majority of the total VCG payment to the agents (particularly for a great number of agents).

\begin{table}[H]
\centering
\begin{tabular}{cccc}
\hline
MLP+FEED & Uniform(0,1) & Normal (0.5,0.1) & n-1 \\ \hline
n=3      & 2.079        & 2.101    & 2        \\
n=4      & 3.071        & 3.111    & 3       \\
n=5      & 4.061        & 4.142    & 4      \\
n=6      & 5.027        & 5.034    & 5       \\
n=7      & 6.009        & 6.067    & 6       \\
n=8      & 7.008        & 7.023    & 7       \\
n=9      & 8.002        & 8.008    & 8       \\
n=10     & 9.003        & 9.023    & 9       \\ \hline
\end{tabular}
\caption{$\sum_{i}{h(\theta_{-i})} / S(\theta)$ in Expectation for Different Distributions}
\label{tab:Expectation}
\end{table}


\section{Conclusion}

In this paper, we consider designing optimal redistribution mechanisms for the public project problem under two objectives: worst-case optimal and optimal-in-expectation. With effective technical improvements on existing networks, we train a neural network to design good redistribution functions. We use a GAN network to generate valuation profiles to find the worst case, and feed prior distribution into loss function to get quality gradients for the optimal-in-expectation objective. To deal with large numbers of agents, we study different dimension-reducing methods and supervise the network into the existing manual mechanism as initialization.
Our experiments show that for the worst case, we could find better worst-case mechanisms compared to existing mechanisms, and for expectation, the neural networks can derive near-optimal redistribution mechanisms.

\bibliographystyle{ACM-Reference-Format}
\bibliography{reference}


\begin{thebibliography}{24}


\ifx \showCODEN    \undefined \def \showCODEN     #1{\unskip}     \fi
\ifx \showDOI      \undefined \def \showDOI       #1{#1}\fi
\ifx \showISBNx    \undefined \def \showISBNx     #1{\unskip}     \fi
\ifx \showISBNxiii \undefined \def \showISBNxiii  #1{\unskip}     \fi
\ifx \showISSN     \undefined \def \showISSN      #1{\unskip}     \fi
\ifx \showLCCN     \undefined \def \showLCCN      #1{\unskip}     \fi
\ifx \shownote     \undefined \def \shownote      #1{#1}          \fi
\ifx \showarticletitle \undefined \def \showarticletitle #1{#1}   \fi
\ifx \showURL      \undefined \def \showURL       {\relax}        \fi
\providecommand\bibfield[2]{#2}
\providecommand\bibinfo[2]{#2}
\providecommand\natexlab[1]{#1}
\providecommand\showeprint[2][]{arXiv:#2}

\bibitem[\protect\citeauthoryear{Arora, Basu, Mianjy, and Mukherjee}{Arora
  et~al\mbox{.}}{2016}]%
        {arora2016understanding}
\bibfield{author}{\bibinfo{person}{Raman Arora}, \bibinfo{person}{Amitabh
  Basu}, \bibinfo{person}{Poorya Mianjy}, {and} \bibinfo{person}{Anirbit
  Mukherjee}.} \bibinfo{year}{2016}\natexlab{}.
\newblock \showarticletitle{Understanding deep neural networks with rectified
  linear units}.
\newblock \bibinfo{journal}{\emph{arXiv preprint arXiv:1611.01491}}
  (\bibinfo{year}{2016}).
\newblock


\bibitem[\protect\citeauthoryear{Cavallo}{Cavallo}{2006}]%
        {Cavallo2006:Optimal}
\bibfield{author}{\bibinfo{person}{Ruggiero Cavallo}.}
  \bibinfo{year}{2006}\natexlab{}.
\newblock \showarticletitle{Optimal Decision-making with Minimal Waste:
  Strategyproof Redistribution of VCG Payments}. In
  \bibinfo{booktitle}{\emph{Proceedings of the Fifth International Joint
  Conference on Autonomous Agents and Multiagent Systems}} (Hakodate, Japan)
  \emph{(\bibinfo{series}{AAMAS '06})}. \bibinfo{publisher}{ACM},
  \bibinfo{address}{New York, NY, USA}, \bibinfo{pages}{882--889}.
\newblock
\showISBNx{1-59593-303-4}
\urldef\tempurl%
\url{https://doi.org/10.1145/1160633.1160790}
\showDOI{\tempurl}


\bibitem[\protect\citeauthoryear{de~Clippel, Naroditskiy, Polukarov, Greenwald,
  and Jennings}{de~Clippel et~al\mbox{.}}{2014}]%
        {Clippel2014:Destroy}
\bibfield{author}{\bibinfo{person}{Geoffroy de Clippel},
  \bibinfo{person}{Victor Naroditskiy}, \bibinfo{person}{Maria Polukarov},
  \bibinfo{person}{Amy Greenwald}, {and} \bibinfo{person}{Nicholas~R.
  Jennings}.} \bibinfo{year}{2014}\natexlab{}.
\newblock \showarticletitle{Destroy to save}.
\newblock \bibinfo{journal}{\emph{Games and Economic Behavior}}
  \bibinfo{volume}{86} (\bibinfo{year}{2014}), \bibinfo{pages}{392--404}.
\newblock


\bibitem[\protect\citeauthoryear{Duetting, Feng, Narasimhan, Parkes, and
  Ravindranath}{Duetting et~al\mbox{.}}{2019}]%
        {Duetting2019:Optimal}
\bibfield{author}{\bibinfo{person}{Paul Duetting}, \bibinfo{person}{Zhe Feng},
  \bibinfo{person}{Harikrishna Narasimhan}, \bibinfo{person}{David Parkes},
  {and} \bibinfo{person}{Sai~Srivatsa Ravindranath}.}
  \bibinfo{year}{2019}\natexlab{}.
\newblock \showarticletitle{Optimal Auctions through Deep Learning}. In
  \bibinfo{booktitle}{\emph{Proceedings of the 36th International Conference on
  Machine Learning}} \emph{(\bibinfo{series}{Proceedings of Machine Learning
  Research}, Vol.~\bibinfo{volume}{97})},
  \bibfield{editor}{\bibinfo{person}{Kamalika Chaudhuri} {and}
  \bibinfo{person}{Ruslan Salakhutdinov}} (Eds.). \bibinfo{publisher}{PMLR},
  \bibinfo{address}{Long Beach, California, USA}, \bibinfo{pages}{1706--1715}.
\newblock


\bibitem[\protect\citeauthoryear{Faltings}{Faltings}{2005}]%
        {Faltings2005:Budget-Balanced}
\bibfield{author}{\bibinfo{person}{Boi Faltings}.}
  \bibinfo{year}{2005}\natexlab{}.
\newblock \showarticletitle{A Budget-Balanced, Incentive-Compatible Scheme for
  Social Choice}. In \bibinfo{booktitle}{\emph{Lecture notes in computer
  science}}. \bibinfo{publisher}{Springer}, \bibinfo{address}{Berlin},
  \bibinfo{pages}{30--43}.
\newblock
\showISBNx{9783540297376}
\showISSN{0302-9743}


\bibitem[\protect\citeauthoryear{Golowich, Narasimhan, and Parkes}{Golowich
  et~al\mbox{.}}{2018}]%
        {Golowich2018:Deep}
\bibfield{author}{\bibinfo{person}{Noah Golowich}, \bibinfo{person}{Harikrishna
  Narasimhan}, {and} \bibinfo{person}{David~C. Parkes}.}
  \bibinfo{year}{2018}\natexlab{}.
\newblock \showarticletitle{Deep Learning for Multi-Facility Location Mechanism
  Design}. In \bibinfo{booktitle}{\emph{Proceedings of the Twenty-Seventh
  International Joint Conference on Artificial Intelligence, {IJCAI-18}}}.
  \bibinfo{publisher}{International Joint Conferences on Artificial
  Intelligence Organization}, \bibinfo{pages}{261--267}.
\newblock


\bibitem[\protect\citeauthoryear{Gujar and Narahari}{Gujar and
  Narahari}{2011}]%
        {Gujar2011:Redistribution}
\bibfield{author}{\bibinfo{person}{Sujit Gujar} {and} \bibinfo{person}{Y.
  Narahari}.} \bibinfo{year}{2011}\natexlab{}.
\newblock \showarticletitle{Redistribution Mechanisms for Assignment of
  Heterogeneous Objects}.
\newblock \bibinfo{journal}{\emph{J. Artif. Intell. Res.}}
  \bibinfo{volume}{41} (\bibinfo{year}{2011}), \bibinfo{pages}{131--154}.
\newblock


\bibitem[\protect\citeauthoryear{Guo}{Guo}{2011}]%
        {Guo2011:VCG}
\bibfield{author}{\bibinfo{person}{Mingyu Guo}.}
  \bibinfo{year}{2011}\natexlab{}.
\newblock \showarticletitle{{VCG} Redistribution with Gross Substitutes}. In
  \bibinfo{booktitle}{\emph{Proceedings of the Twenty-Fifth {AAAI} Conference
  on Artificial Intelligence, {AAAI} 2011, San Francisco, California, USA,
  August 7-11, 2011}}, \bibfield{editor}{\bibinfo{person}{Wolfram Burgard}
  {and} \bibinfo{person}{Dan Roth}} (Eds.). \bibinfo{publisher}{{AAAI} Press}.
\newblock
\urldef\tempurl%
\url{http://www.aaai.org/ocs/index.php/AAAI/AAAI11/paper/view/3733}
\showURL{%
\tempurl}


\bibitem[\protect\citeauthoryear{Guo}{Guo}{2012}]%
        {Guo2012:Worst}
\bibfield{author}{\bibinfo{person}{Mingyu Guo}.}
  \bibinfo{year}{2012}\natexlab{}.
\newblock \showarticletitle{Worst-case optimal redistribution of {VCG} payments
  in heterogeneous-item auctions with unit demand}. In
  \bibinfo{booktitle}{\emph{International Conference on Autonomous Agents and
  Multiagent Systems, {AAMAS} 2012, Valencia, Spain, June 4-8, 2012 {(3}
  Volumes)}}, \bibfield{editor}{\bibinfo{person}{Wiebe van~der Hoek},
  \bibinfo{person}{Lin Padgham}, \bibinfo{person}{Vincent Conitzer}, {and}
  \bibinfo{person}{Michael Winikoff}} (Eds.). \bibinfo{publisher}{{IFAAMAS}},
  \bibinfo{pages}{745--752}.
\newblock
\urldef\tempurl%
\url{http://dl.acm.org/citation.cfm?id=2343803}
\showURL{%
\tempurl}


\bibitem[\protect\citeauthoryear{Guo}{Guo}{2016}]%
        {Guo2016:Competitive}
\bibfield{author}{\bibinfo{person}{Mingyu Guo}.}
  \bibinfo{year}{2016}\natexlab{}.
\newblock \showarticletitle{Competitive {VCG} Redistribution Mechanism for
  Public Project Problem}. In \bibinfo{booktitle}{\emph{{PRIMA} 2016: Princiles
  and Practice of Multi-Agent Systems - 19th International Conference, Phuket,
  Thailand, August 22-26, 2016, Proceedings}} \emph{(\bibinfo{series}{Lecture
  Notes in Computer Science}, Vol.~\bibinfo{volume}{9862})}.
  \bibinfo{publisher}{Springer}, \bibinfo{pages}{279--294}.
\newblock


\bibitem[\protect\citeauthoryear{Guo}{Guo}{2019}]%
        {Guo2019:Asymptotically}
\bibfield{author}{\bibinfo{person}{Mingyu Guo}.}
  \bibinfo{year}{2019}\natexlab{}.
\newblock \showarticletitle{An Asymptotically Optimal VCG Redistribution
  Mechanism for the Public Project Problem}. In
  \bibinfo{booktitle}{\emph{Proceedings of the Twenty-Eighth International
  Joint Conference on Artificial Intelligence, {IJCAI-19}}}.
  \bibinfo{publisher}{International Joint Conferences on Artificial
  Intelligence Organization}, \bibinfo{pages}{315--321}.
\newblock


\bibitem[\protect\citeauthoryear{Guo and Conitzer}{Guo and Conitzer}{2009}]%
        {Guo2009:Worst}
\bibfield{author}{\bibinfo{person}{Mingyu Guo} {and} \bibinfo{person}{Vincent
  Conitzer}.} \bibinfo{year}{2009}\natexlab{}.
\newblock \showarticletitle{Worst-case optimal redistribution of {VCG} payments
  in multi-unit auctions}.
\newblock \bibinfo{journal}{\emph{Games and Economic Behavior}}
  \bibinfo{volume}{67}, \bibinfo{number}{1} (\bibinfo{year}{2009}),
  \bibinfo{pages}{69--98}.
\newblock


\bibitem[\protect\citeauthoryear{Guo and Conitzer}{Guo and Conitzer}{2014}]%
        {Guo2014:Better}
\bibfield{author}{\bibinfo{person}{Mingyu Guo} {and} \bibinfo{person}{Vincent
  Conitzer}.} \bibinfo{year}{2014}\natexlab{}.
\newblock \showarticletitle{Better redistribution with inefficient allocation
  in multi-unit auctions}.
\newblock \bibinfo{journal}{\emph{Artificial Intelligence}}
  \bibinfo{volume}{216} (\bibinfo{year}{2014}), \bibinfo{pages}{287--308}.
\newblock


\bibitem[\protect\citeauthoryear{Guo and Shen}{Guo and Shen}{2017}]%
        {Guo2017:Speed}
\bibfield{author}{\bibinfo{person}{Mingyu Guo} {and} \bibinfo{person}{Hong
  Shen}.} \bibinfo{year}{2017}\natexlab{}.
\newblock \showarticletitle{Speed up Automated Mechanism Design by Sampling
  Worst-Case Profiles: An Application to Competitive {VCG} Redistribution
  Mechanism for Public Project Problem}. In \bibinfo{booktitle}{\emph{{PRIMA}
  2017: Principles and Practice of Multi-Agent Systems - 20th International
  Conference, Nice, France, October 30 - November 3, 2017, Proceedings}}
  \emph{(\bibinfo{series}{Lecture Notes in Computer Science},
  Vol.~\bibinfo{volume}{10621})}. \bibinfo{publisher}{Springer},
  \bibinfo{pages}{127--142}.
\newblock


\bibitem[\protect\citeauthoryear{Keskar, Mudigere, Nocedal, Smelyanskiy, and
  Tang}{Keskar et~al\mbox{.}}{2016}]%
        {keskar2017large}
\bibfield{author}{\bibinfo{person}{Nitish~Shirish Keskar},
  \bibinfo{person}{Dheevatsa Mudigere}, \bibinfo{person}{Jorge Nocedal},
  \bibinfo{person}{Mikhail Smelyanskiy}, {and} \bibinfo{person}{Ping Tak~Peter
  Tang}.} \bibinfo{year}{2016}\natexlab{}.
\newblock \showarticletitle{On large-batch training for deep learning:
  Generalization gap and sharp minima}.
\newblock \bibinfo{journal}{\emph{ICLR 2017}} (\bibinfo{year}{2016}).
\newblock


\bibitem[\protect\citeauthoryear{Manisha, Jawahar, and Gujar}{Manisha
  et~al\mbox{.}}{2018}]%
        {Manisha2018:Learning}
\bibfield{author}{\bibinfo{person}{Padala Manisha}, \bibinfo{person}{C.~V.
  Jawahar}, {and} \bibinfo{person}{Sujit Gujar}.}
  \bibinfo{year}{2018}\natexlab{}.
\newblock \showarticletitle{Learning Optimal Redistribution Mechanisms Through
  Neural Networks}. In \bibinfo{booktitle}{\emph{Proceedings of the 17th
  International Conference on Autonomous Agents and MultiAgent Systems, {AAMAS}
  2018, Stockholm, Sweden, July 10-15, 2018}},
  \bibfield{editor}{\bibinfo{person}{Elisabeth Andr{\'{e}}},
  \bibinfo{person}{Sven Koenig}, \bibinfo{person}{Mehdi Dastani}, {and}
  \bibinfo{person}{Gita Sukthankar}} (Eds.). \bibinfo{publisher}{International
  Foundation for Autonomous Agents and Multiagent Systems Richland, SC, {USA} /
  {ACM}}, \bibinfo{pages}{345--353}.
\newblock


\bibitem[\protect\citeauthoryear{Mas-Colell, Whinston, and Green}{Mas-Colell
  et~al\mbox{.}}{1995}]%
        {Mas-Colell1995:Microeconomic}
\bibfield{author}{\bibinfo{person}{Andreu Mas-Colell}, \bibinfo{person}{Michael
  Whinston}, {and} \bibinfo{person}{Jerry~R. Green}.}
  \bibinfo{year}{1995}\natexlab{}.
\newblock \bibinfo{booktitle}{\emph{Microeconomic Theory}}.
\newblock \bibinfo{publisher}{Oxford University Press}.
\newblock


\bibitem[\protect\citeauthoryear{Moore}{Moore}{2006}]%
        {Moore2006:General}
\bibfield{author}{\bibinfo{person}{J. Moore}.} \bibinfo{year}{2006}\natexlab{}.
\newblock \bibinfo{booktitle}{\emph{General Equilibrium and Welfare Economics:
  An Introduction}}.
\newblock \bibinfo{publisher}{Springer}.
\newblock


\bibitem[\protect\citeauthoryear{Moulin}{Moulin}{1988}]%
        {Moulin1988:Axioms}
\bibfield{author}{\bibinfo{person}{H. Moulin}.}
  \bibinfo{year}{1988}\natexlab{}.
\newblock \bibinfo{booktitle}{\emph{Axioms of Cooperative Decision Making}}.
\newblock \bibinfo{publisher}{Cambridge University Press}.
\newblock


\bibitem[\protect\citeauthoryear{Moulin}{Moulin}{2009}]%
        {Moulin2009:Almost}
\bibfield{author}{\bibinfo{person}{Herv\'e Moulin}.}
  \bibinfo{year}{2009}\natexlab{}.
\newblock \showarticletitle{Almost budget-balanced {VCG} mechanisms to assign
  multiple objects}.
\newblock \bibinfo{journal}{\emph{JET}} \bibinfo{volume}{144},
  \bibinfo{number}{1} (\bibinfo{year}{2009}), \bibinfo{pages}{96--119}.
\newblock


\bibitem[\protect\citeauthoryear{Naroditskiy, Guo, Dufton, Polukarov, and
  Jennings}{Naroditskiy et~al\mbox{.}}{2012}]%
        {Naroditskiy2012:Redistribution}
\bibfield{author}{\bibinfo{person}{Victor Naroditskiy}, \bibinfo{person}{Mingyu
  Guo}, \bibinfo{person}{Lachlan Dufton}, \bibinfo{person}{Maria Polukarov},
  {and} \bibinfo{person}{Nicholas~R. Jennings}.}
  \bibinfo{year}{2012}\natexlab{}.
\newblock \showarticletitle{Redistribution of {VCG} Payments in Public Project
  Problems}. In \bibinfo{booktitle}{\emph{Internet and Network Economics - 8th
  International Workshop, {WINE} 2012, Liverpool, UK, December 10-12, 2012.
  Proceedings}} \emph{(\bibinfo{series}{Lecture Notes in Computer Science},
  Vol.~\bibinfo{volume}{7695})}, \bibfield{editor}{\bibinfo{person}{Paul~W.
  Goldberg}} (Ed.). \bibinfo{publisher}{Springer}, \bibinfo{pages}{323--336}.
\newblock


\bibitem[\protect\citeauthoryear{Schulman, Heess, Weber, and Abbeel}{Schulman
  et~al\mbox{.}}{2015}]%
        {schulman2015gradient}
\bibfield{author}{\bibinfo{person}{John Schulman}, \bibinfo{person}{Nicolas
  Heess}, \bibinfo{person}{Theophane Weber}, {and} \bibinfo{person}{Pieter
  Abbeel}.} \bibinfo{year}{2015}\natexlab{}.
\newblock \showarticletitle{Gradient estimation using stochastic computation
  graphs}.
\newblock \bibinfo{journal}{\emph{arXiv preprint arXiv:1506.05254}}
  (\bibinfo{year}{2015}).
\newblock


\bibitem[\protect\citeauthoryear{Shen, Tang, and Zuo}{Shen
  et~al\mbox{.}}{2019}]%
        {Shen2019:Automated}
\bibfield{author}{\bibinfo{person}{Weiran Shen}, \bibinfo{person}{Pingzhong
  Tang}, {and} \bibinfo{person}{Song Zuo}.} \bibinfo{year}{2019}\natexlab{}.
\newblock \showarticletitle{Automated Mechanism Design via Neural Networks}. In
  \bibinfo{booktitle}{\emph{Proceedings of the 18th International Conference on
  Autonomous Agents and MultiAgent Systems}} (Montreal QC, Canada)
  \emph{(\bibinfo{series}{AAMAS '19})}. \bibinfo{publisher}{International
  Foundation for Autonomous Agents and Multiagent Systems},
  \bibinfo{address}{Richland, SC}, \bibinfo{pages}{215--223}.
\newblock
\showISBNx{978-1-4503-6309-9}


\bibitem[\protect\citeauthoryear{Wang, Guo, Sakurai, Babar, and Guo}{Wang
  et~al\mbox{.}}{2021}]%
        {Wang2021:Mechanism}
\bibfield{author}{\bibinfo{person}{Guanhua Wang}, \bibinfo{person}{Runqi Guo},
  \bibinfo{person}{Yuko Sakurai}, \bibinfo{person}{Ali Babar}, {and}
  \bibinfo{person}{Mingyu Guo}.} \bibinfo{year}{May 2021}\natexlab{}.
\newblock \showarticletitle{Mechanism Design for Public Projects via Neural
  Networks}. In \bibinfo{booktitle}{\emph{20th International Conference on
  Autonomous Agents and Multiagent Systems ({AAMAS} 2021, online)}}.
\newblock


\end{thebibliography}






\end{document}